\documentclass[12pt]{article}
\usepackage{cite}
\usepackage[dvips]{psfrag}
\usepackage{amsmath}
\usepackage{graphicx}

\begin{document}

\setlength{\textheight}{21.5cm}
\setlength{\oddsidemargin}{0.cm}
\setlength{\evensidemargin}{0.cm}
\setlength{\topmargin}{0.cm}
\setlength{\footskip}{1cm}
\setlength{\arraycolsep}{2pt}

\renewcommand{\thefootnote}{\#\arabic{footnote}}
\setcounter{footnote}{0}

\newcommand{\gtrsim}{ \mathop{}_{\textstyle \sim}^{\textstyle >} }
\newcommand{\lesssim}{ \mathop{}_{\textstyle \sim}^{\textstyle <} }
\newcommand{\rem}[1]{{\bf #1}}
\renewcommand{\thefootnote}{\fnsymbol{footnote}}
\setcounter{footnote}{0}
\begin{titlepage}
\def\thefootnote{\fnsymbol{footnote}}

\begin{center}
\hfill 
January 2008\\
\vskip .5in
\bigskip
\bigskip
{\Large \bf Constraints on Deflation from the Equation of State
of Dark Energy}

\vskip .45in

{\bf Lauris Baum, Paul H. Frampton and Shinya Matsuzaki}

\vskip .45in

{\it Department of Physics and Astronomy, University of North Carolina,\\
Chapel Hill, NC 27599.}

\end{center}

\vskip .4in
\begin{abstract}
In cyclic cosmology based on phantom dark energy the requirement
that our universe satisfy a CBE-condition ({\it Comes Back Empty})
imposes a lower bound on the number $N_{\rm cp}$ of causal patches
which separate just prior to turnaround. This bound depends on the dark
energy equation of state $w = p/\rho = -1 - \phi$ with $\phi > 0$.
More accurate measurement of $\phi$ will constrain
$N_{\rm cp}$. The critical density $\rho_c$ in the model has a lower
bound $\rho_c \ge (10^9 {\rm GeV})^4$ or $\rho_c \ge (10^{18} {\rm GeV})^4$
when the smallest bound state has size $10^{-15}$m, or
$10^{-35}$m, respectively. 
\end{abstract}
\end{titlepage}

\renewcommand{\thepage}{\arabic{page}}
\setcounter{page}{1}
\renewcommand{\thefootnote}{\#\arabic{footnote}}

\newpage

\section{Introduction}

\bigskip

Recently two of the authors have proposed~\cite{BF,F,BF2}
a scenario for a cyclic universe based on a dark
energy component with constant equation of state
satisfying $w = p/\rho = -1-\phi$ where $w < -1$
and hence $\phi > 0$. The model involves two key
ideas: (i) that the universe deflate just prior to the turnaround
from expansion to contraction by disintegrating
into a very large number $N_{\rm cp}$ of causal
patches. In the notation of \cite{BF},
note that $N_{\rm cp}=1/f^3$; (ii) that the contracting
universe be empty, meaning that one causal
patch at turnaround must contain no matter or
black holes, only dark energy. This is called the CBE condition
({\it Comes Back Empty}). Implementation
of CBE requires, as we shall explain, a lower 
bound on $N_{cp}$ which depends on the length
scale $L$ characterising the smallest bound system.
We shall consider both $L = 10^{-15}$m for a nucleon
then $L \ge 10^{-35}$m for a PPP ({\it Presently Point
Particle}) meaning a particle
which at present is considered to be pointlike,
like a quark or a lepton, 
but which actually has a characteristic size 
at least a few orders of magnitude smaller than a nucleon
but greater than or equal to the Planck scale.

\bigskip

In the foreseeable future, it is expected that the equation
of state of the dark energy $w$, and hence $\phi$, 
will be measured with higher accuracy by, for example,
the Planck Surveyor satellite~\cite{Planck}. What we shall show
is that this measurement can, within this model,
constrain for a given $w$ the number $N_{\rm cp}$
of causal patches at turnaround by imposing
a lower bound thereon. 

\bigskip

The plan of the paper is that in Section 2 we discuss the times at which unbinding,
causal disconnection and turnaround occur.
In Section 3, the constraints on $N_{\rm cp}$ from measurement of $\phi$
are derived. Finally, Section 4 is a discussion.
In the Appendices is technical material
to supplement the main text.

\newpage

\section{Times of unbinding, causal disconnection and turnaround}

In this section 
we analyze four relationships between cosmic times in the cyclic model expansion era: 
i)  $t_{\rm unbound}$ (at which a bound system will become unbound due to the large dark energy 
force with $w<-1$); 
ii) $t_{\rm caus}$ (at which a previously bound system becomes casually disconnected, meaning
that no light signal could exchange before the would-be Big Rip; this is how
we estimate $N_{\rm cp}$); 
iii) $t_T$ (time when the turnaround occurs);  
iv) $t_{\rm rip}$ (at which a ``would-be" big rip takes place),  
in addition to the present time $t_0$.

In the Baum-Frampton (BF) model~\cite{BF},  
there are three parameters;  
$w$ (equation of state of dark energy); 
$\rho_C$ (critical density which the total density in the system $\rho_{\rm tot}$ reached at $t=t_T$); 
$f$ (the deflation fraction parameter related to the number of causal patches by $N_{\rm cp}=(1/f^3)$).  
We will analyze the model  
taking the value of $w$ lying in a range, 
\begin{equation} 
  -1.10000 \le \omega \le -1.00001
  \,, 
\end{equation}
and for $\rho_C$ choosing the following range, 
\begin{equation} 
(10^3 \, {\rm GeV})^4 \le \rho_C \le (10^{19} \, {\rm GeV})^4   
\,.         
\end{equation}
The choice of the range of $w$ is motivated by the current lower bound from  
observations~\cite{Yao:2006px,WMAP3} and the upper bound, 
by the cosmic variance uncertainty in this measurement.

  Reserving details of the derivation of four formulas to Appendix A, 
  we shall here refer to the resultant expressions:

\begin{itemize} 

\item 
$(t_{\rm rip}-t_0)$ 

\begin{equation} 
  t_{\rm rip} - t_0 \simeq \frac{11\, {\rm Gyr}}{|1+w|} 
  \,. \label{tript0:eq}
\end{equation}

\item 

$(t_{\rm rip}-t_{\rm unbound})$

\begin{equation} 
  t_{\rm rip} - t_{\rm unbound} = \alpha(w) P 
  \,, \label{triptu:eq}
\end{equation}
where~\cite{Caldwell:2003vq}  
\begin{equation} 
 \alpha (w) = \frac{\sqrt{2 |1+3w|}}{6\pi |1+w|} 
 \,, 
\end{equation}
  and $P$ denotes the period associated with the binding force 
  which had been constraining objects into a certain bound system 
before $t=t_{\rm unbound}$.

\item 

$(t_{\rm rip} - t_{\rm caus})$

\begin{equation} 
  t_{\rm rip} - t_{\rm caus} 
= \Bigg|  \frac{1+3w}{3(1+w)}  \Bigg| \left(  \frac{L}{c} \right)
\label{triptcaus}
\end{equation}
where $c$ is the speed of light and $L$ stands for the 
length scale of the bound system~\cite{Caldwell:2003vq}.

\item 

$(t_{\rm rip}-t_T)$

\begin{equation} 
  t_{\rm rip} - t_T = \frac{11\, {\rm Gyr}}{|1+w|}   10^{-14.5} \eta^{-1/2} 
\label{trip}
\end{equation} 
  where $\eta$ is a scale factor of $\rho_C$ defined by $\rho_C = \eta \rho_{H_2O}$ with 
  $\rho_{H_2O}$ being the density if water, $\rho_{H_2O} = 1 {\rm g} \cdot {\rm cm}^{-3}$.  
Eq.(\ref{trip}) appeared 
\footnote{Eq.(\ref{trip}) corrects
a typo in the exponent of $\eta$ appearing in \cite{BF}}
as Eq.(4) in Ref.~\cite{BF}. 
\end{itemize}   

The numerical analysis for these relationships is presented in Appendices A and B. 
As a result, we find the lower bound for $\rho_C$, 
\begin{equation} 
      \rho_C \gtrsim (10^{18} \, {\rm GeV})^4, 
\end{equation}
which is obtained by imposing that the time for a presently point particle (PPP), 
with the size $10^{-33}$m = L,  satisfy  
$t_{\rm rip} > t_T > t_{\rm caus}^{\rm PPP} > t_{\rm unbound}^{\rm PPP}$. 
It should be emphasized that this result is almost independent of 
a choice of $w$ in the range of interest. 
 
For a nucleon with $L \simeq 10^{-15}$m, the corresponding lower bound is
\begin{equation} 
      \rho_C \gtrsim (10^{9} \, {\rm GeV})^4 
\,. 
\end{equation}

\newpage

\section{Given $w$, the constraints on $N_{\rm cp}$}

\bigskip

In \cite{BF}, as in \cite{Caldwell:2003vq}, various bound systems
were discussed including galaxies, the Earth-Sun system, the hydrogen
atom and a nucleon. Each may be charaterised by a present length scale $L_0$.

\bigskip

For the CBE condition we must insist that the smallest bound systems
are disintegrated before turnaround which means that the size of a generic
causal patch $L_{\rm cp}$ (to be defined below)
is smaller than the size $L(t_T)$ at turnaround of
the bound system whose present length scale is $L(t_0) = L_0$, namely

\begin{equation}
L_{\rm cp} \leq L(t_T) = L_0 \left(\frac{a(t_T)}{a(t_{\rm unbound})} \right).
\label{CBE}
\end{equation}

\bigskip

We remind the reader that the CBE condition is mandatory because
if the contracting universe contains matter it will not generally
contract sufficiently but will undergo a premature bounce. Even
if a causal patch contains only one very infra-red photon, this 
can blue-shift to an energy sufficient to create $e^+e^-$ pairs
before the bounce, again disallowing sufficient contraction for 
infinite cyclicity.

\bigskip

This was the motivation for \cite{BF2} where it was shown that
the mean number of low-energy photons per causal 
patch is much less than one and is essentially zero.
There will always be a vanishing but strictly non-zero
number of patches which fail to cycle but it was shown
in \cite{F} that the probability of a successful universe
is equal to one; it was noted that the total number of
universes has always been, and always will be constantly
infinite and equal to $\aleph_0$ (Aleph-zero). $\aleph_0$ is a countable 
infinity, exemplified by the number of primes, of integers or
of rational numbers.

\bigskip

To enable infinite cyclicity we must have the CBE condition,
Eq.(\ref{CBE}), for the smallest bound systems. The smallest
bound systems we know about are nucleons with $L_0 = 10^{-15}$m. 

\bigskip

To be general, we consider PPPs ({\it Presently Point
Particles}) meaning particles
which are presently regarded as pointlike but may not be. 
We allow a bound state 
scale for PPPs to be anywhere between the present upper limit
of about $(1 {\rm TeV})^{-1} = 10^{-18}$m and the Planck scale of
$10^{-35}$m. As we shall see shortly, the lower bound
on $N_{\rm cp}$ is so sensitive to where $L_0$ 
is chosen within these twenty orders of magnitude that
its presentation requires us to plot $\log_{10}\log_{10}N_{\rm cp}$ 
against the equation of state of the dark energy.

\newpage

\begin{figure}

\begin{center}
\vspace{15pt}
\psfrag{w}{\vspace{40pt} \hspace{-15pt} $w$}
\psfrag{loglogncp}{\vspace{-20pt} \hspace{-25pt} $\log_{10}\log_{10}N_{\rm cp}$} 
\vspace{15pt} 
\includegraphics[scale=0.6]{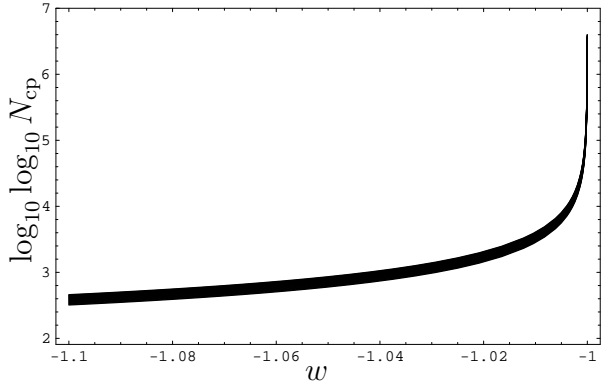} 
\vspace{5pt}
\caption{\footnotesize Constraint on $w$ and $N_{\rm cp}$ coming from the 
CBE condition (Comes Back Empty), corresponding to inequality 
(\ref{cbe}). 
The black band in this figure has been created
by varying the length of smaller bound systems
from $L_p(t_0) = 10^{-33}$m to $L_p(t_0) = 10^{-15}$m; the bottom edge 
corresponds to the lower value of $L_p(t_0)$. 
The region below the black band is forbidden by
the CBE condition. 
}  
\label{cons-w-cbe} 
\end{center}

\end{figure}

\begin{figure}

\begin{center}
\vspace{15pt}
\psfrag{w}{\vspace{40pt} \hspace{-15pt} $w$}
\psfrag{loglogncp}{\vspace{-20pt} \hspace{-25pt} $\log_{10}\log_{10}N_{\rm cp}$} 
\vspace{15pt} 
\includegraphics[scale=0.6]{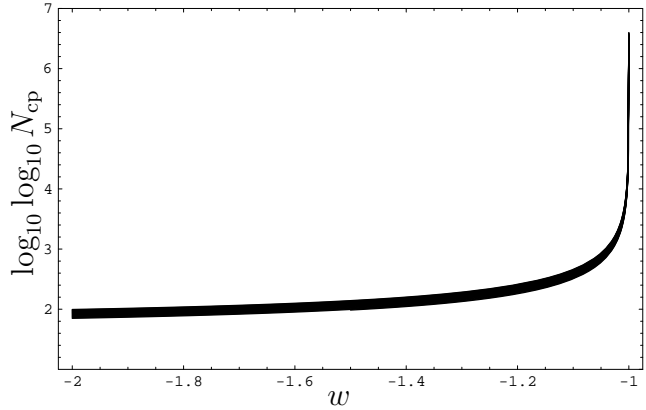} 
\vspace{5pt}
\caption{\footnotesize Constraint for $-2 \leq w \leq -1.1$ and $N_{\rm cp}$ coming from the 
CBE condition (Comes Back Empty), corresponding to inequality 
(\ref{cbe}). 
The black band in this figure has been created
by varying the length of smaller bound systems
from $L_p(t_0) = 10^{-33}$m to $L_p(t_0) = 10^{-15}$m; the bottom edge 
corresponds to the lower value of $L_p(t_0)$. 
The region below the black band is forbidden by
the CBE condition. 
}  
\label{cons-w-extend-cbe} 
\end{center}

\end{figure}

\bigskip
\bigskip

The present Hubble length $r_H(t_0)$ is given by
\begin{equation}
r_H(t_0) = \frac{1}{H_0}
\label{hubblelength}
\end{equation}
which, at the turnaround, would naively become
\begin{equation}
r_H(t_T) = r_H(t_0) a(t_T)
\end{equation}
since by definition $a(t_0) = 1$.

\bigskip

In the cyclic model of \cite{BF}, the size of a causal
patch $L_{\rm cp}$ is instead defined by
\begin{equation} 
  L_{\rm cp} = \frac{r_H(t_T)}{N_{\rm cp}}  
\label{cbe}
\end{equation}
and therefore Eq.(\ref{CBE}) can be calculated for different values
of $L_0$, see Appendix C. The results are illustrated in Figure 1
where we plot $\log_{10}\log_{10}N_{\rm cp}$ versus $w = -1 -\phi$.

\bigskip

From this figure we see that a measurement of $w$ in the range
anticipated for the Planck surveyor
will provide a lower bound on $N_{\rm cp}$. For example $w=-1.05$ implies
$N_{\rm cp} \gtrsim 10^{630}$ for disintegration of nucleons and
$N_{\rm cp} \gtrsim 10^{1000}$ for disintegration of PPPs with bound scale at
the Planck length.

\bigskip

Since we know the entropy of the present universe is at least
$S(t_0) \gtrsim 10^{102}$ \cite{FK,FHKR} one must impose
\begin{equation}
N_{\rm cp} \gtrsim 10^{102}
\label{102}
\end{equation}
and, by requiring only the dissociation of nucleons we see from
Figure 2 that this implies 

\begin{equation}
w \gtrsim -2.
\label{w2}
\end{equation}

\bigskip

 Of course, WMAP data~\cite{WMAP3} 
already guarantee this condition but it is interesting that the
cyclic model of \cite{BF} would be impossible if Eq.(\ref{w2})
had been violated.

\newpage

\section{Discussion}

\bigskip

What we have deduced is that the parameters $\rho_c, w$ and $N_{\rm cp}$
in cyclic cosmology are already constrained by existing data. For
example one requires $w \gtrsim -2$ for the CBE aspect to work. 

\bigskip

This constraint is already known to be respected in Nature but
as better and more accurate cosmological data become available
it will shed further light on the viability of the theory.

\bigskip

In particular, the accurate measurement of the equation of state
$w = -1 -\phi$ is of special interest. Fortunately the Planck Surveyor~\cite{Planck} 
is anticipated to acquire improved accuracy on $w$ in the near future.
As we have discussed, this will provide a lower bound on the number
$N_{\rm cp}$ of causal patches necessary to dissociate the smallest bound
systems at turnaround and hence to solve the entropy problem
and, via CBE, enable the possibility of infinite cyclicity.

\bigskip

It is amusing that the physical conditions at the approach of deflation 
are so extraordinary that it is natural to ask whether the systems
presently regarded as point particles may
be composite because the phantom dark energy density grows
to unimaginably large values and can disintegrate bound systems
down to arbitrarily small scales. We have conservatively limited our
attention to systems bigger than the Planck length. However, although
this requirement seems dictated by considerations of quantum gravity,
it is possible that the dark energy will dissociate even 
smaller systems if they exist.

\bigskip

The advantage of cyclic cosmology is that it removes the initial singularity
associated with the Big Bang, about 13.7 billion years ago, and allows that time
never began. The previous attempts to create a consistent infinite
cyclicity were stymied between about 1934~\cite{Tolman} and
2002~\cite{ST} primarily because of the entropy problem and the second law
of thermodynamics. The discovery of the accelerated expansion rate
of the universe and the concomitant necessity of dark energy
has permitted more optimism that the cyclic cosmology is, after all,
on the right track. 

\newpage

\begin{center}

\section*{Acknowledgments}
\end{center}

This work was supported in part by the U.S. Department of Energy
under Grant No. DG-FG02-06ER41418.

\newpage 

\Large 

\noindent 
{\bf Appendix A}

\bigskip 

\noindent 
{\bf Derivation of formulas and numerical analysis} 

\bigskip 

\normalsize 
We begin by writing down 
the Friedman equation for times
$(t_0<t<t_T$ which correspond to the expansion phase~\cite{BF},  
\begin{equation}
H^2(t) \equiv 
  \left(  \frac{\dot{a}(t)}{a(t)} \right)^2 
  = \frac{8\pi G}{3} \left[ \frac{\rho_\Lambda^0}{[a(t)]^{3(1+w)}} - \frac{[\rho_{\rm tot}(t)]^2 }{\rho_C} \right] 
\,,\label{FR}
\end{equation}
where we have put $\Omega_r^0=\Omega_m^0=0$. 
Taking into account the rapid acceleration when $t_0 < t<t_T$ (or $t_{\rm rip}$), we may neglect the last term 
proportional to $\rho^2_{\rm tot} \sim [a^{3(1+w)}]^2$, so that 
\begin{equation} 
    \left(  \frac{\dot{a}(t)}{a(t)} \right)^2 \simeq H_0^2 
   \frac{\Omega_\Lambda^0}{[a(t)]^{3(1+w)}} 
\,. \label{Feq}
\end{equation}

With the boundary condition $a(t_{\rm rip})=\infty$, 
by employing the equation of state $w <-1$, 
Eq.(\ref{Feq}) can readily be solved for an arbitrary time $t$ satisfying $t<t_{\rm rip}$ to get 
\begin{equation} 
  t_{\rm rip} -t = ( H_0  \sqrt{\Omega_\Lambda^0})^{-1}  \frac{2}{3|1+w|} a(t)^{-\frac{3|1+w|}{2}} 
  \,. \label{tripto}
\end{equation}

\subsection*{A1. The formula for ($t_{\rm rip}-t_0$)}

Taking $t=t_0$ at which point $a(t_0)=1$ and 
using current observational values~\cite{Yao:2006px}, 
$H_0=73 \, {\rm km} \cdot {\rm s}^{-1} \cdot {\rm Mpc}^{-1}$ and 
$\Omega_\Lambda^0 = 0.76$, 
we find the time interval ($t_{\rm rip} - t_0$) from Eq.(\ref{tripto}) to be~\cite{Caldwell:2003vq} 
\begin{equation} 
  t_{\rm rip} - t_0 \simeq \frac{11\, {\rm Gyr}}{|1+w|} 
  \,. 
\end{equation}
In Table~\ref{triptot0} we list 
the values of ($t_{\rm rip} - t_0$) for 37 specific choices of $w$ in the range  
$-1.10000 \le \omega \le -1.00001$.

\begin{table}

\begin{center}

\begin{tabular}{c|r} 
\hline 
\hspace{55pt} $w$  \hspace{55pt} & 
\hspace{55pt} ($t_{\rm rip} - t_0$) [Gyr] \hspace{55pt} 
\\ 
\hline \hline 
$-1.10000$ & 110 \\ 
$-1.09000$ & 122 \\
$-1.08000$ & 137 \\
$-1.07000$ & 157 \\
$-1.06000$ & 183 \\
$-1.05000$ & 220 \\
$-1.04000$ & 275 \\
$-1.03000$ & 366 \\
$-1.02000$ & 550 \\
$-1.01000$ & 1100 \\
$-1.00900$ & 1222 \\
$-1.00800$ & 1375 \\
$-1.00700$ & 1571 \\
$-1.00600$ & 1833 \\
$-1.00500$ & 2200 \\
$-1.00400$ & 2750 \\
$-1.00300$ & 3666 \\
$-1.00200$ & 5500 \\
$-1.00100$ & 11000 \\
$-1.00090$ & 12222 \\
$-1.00080$ & 13750 \\
$-1.00070$ & 15714 \\
$-1.00060$ & 18333 \\
$-1.00050$ & 22000 \\
$-1.00040$ & 27500 \\
$-1.00030$ & 36666 \\
$-1.00020$ & 55000 \\
$-1.00010$ & 110000 \\
$-1.00009$ & 122222 \\
$-1.00008$ & 137500 \\
$-1.00007$ & 157143 \\
$-1.00006$ & 183333 \\
$-1.00005$ & 220000 \\
$-1.00004$ & 275000 \\
$-1.00003$ & 366667 \\
$-1.00002$ & 550000 \\
$-1.00001$ & 1100000 \\
\hline 
\end{tabular} 
\vspace{15pt} 
\caption{ Values of ($t_{\rm rip} - t_0$) for $-1.10000 \le \omega \le -1.00001$.  } 
\label{triptot0}
\end{center}

\end{table}

\newpage

\subsection*{A2. The formula for ($t_{\rm rip}-t_T$)}

Putting $t=t_T$ in Eq.(\ref{tripto}) and dividing both sides by 
$(t_{\rm rip} - t_0 )$, 
we find a relationship independent of both $H_0$ and $\Omega^0_\Lambda$, 
\begin{equation} 
   \frac{t_{\rm rip} - t_T }{t_{\rm rip} - t_0 } = [a(t_T)]^{ - \frac{3\phi}{2}} 
   \,. \label{trip-tT:ratio}
\end{equation}
 We shall recall here that the turnaround-time $t_T$ is characterized by 
 $\rho_\Lambda(t_T)=\rho_C$, derived from examining a solution $H^2=0$ of Eq.(\ref{FR}), 
which allows us to rewrite Eq.(\ref{trip-tT:ratio}) as 
 \begin{equation} 
   \frac{t_{\rm rip} - t_T }{t_{\rm rip} - t_0 } = 
\sqrt{ \frac{\rho_\Lambda^0}{\rho_C}}
   \,, 
\end{equation}
 where we have calculated the right-hand side using 
\begin{equation} 
[a(t_T)]^{-3\phi}=\frac{\rho_\Lambda(t_0)}{\rho_\Lambda(t_T)} 
= \frac{\rho_\Lambda(t_0)}{\rho_C}  
\,. 
\end{equation}

   Following Ref.~\cite{BF}, 
   we may introduce a unit of energy density, $\rho_{H_2O}$, in such a way that 
the critical density $\rho_C$ is scaled by a factor of $\eta$   
\begin{equation} 
  \rho_C \equiv \eta \cdot \rho_{H_2O} 
  \,. 
\end{equation}
The present dark energy density $\rho_\Lambda^0$ can be expressed  
in terms of $\rho_{H_2O} = 1{\rm g/  cm^{3}}$ as 
\begin{equation}
   \rho_\Lambda^0 = 10^{-29} \rho_{H_2O}
   \,, 
\end{equation} 
which immediately leads us to 
\begin{equation} 
\frac{\rho_\Lambda^0}{\rho_C} = \frac{10^{-29} \rho_{H_2O}}{\eta \cdot \rho_{H_2O} } 
  = \eta^{-1}  10^{-29} 
\,. 
\end{equation}
Hence we have~\cite{BF} 
\begin{equation} 
  t_{\rm rip} - t_T = (t_{\rm rip} - t_0 ) \cdot 10^{-14.5} \cdot \eta^{-1/2} 
  \,. \label{triptotT}
\end{equation} 
In Tables~\ref{triptotT29} to \ref{triptotT93}, 
choosing $\eta = 10^{29}, 10^{57}, 10^{93}$, respectively, corresponding to 
$\rho_C \simeq (10^3 \, {\rm GeV})^4$, 
$(10^{10} \, {\rm GeV})^4, (10^{19} \, {\rm GeV})^4$ in units of $\rho_{H_2 O}$, 
we list the values for the time interval $(t_{\rm rip} - t_T)$, 
which turn out to be at most of order ${\cal O}(10^{-7}\, {\rm s})$.

\begin{table} 

\begin{center}

$\eta=10^{29} $ 

\begin{tabular}{c|r} 
\hline 
\hspace{55pt} $w$  \hspace{55pt} & 
\hspace{55pt} $(t_{\rm rip} - t_T) $ [s] \hspace{55pt} 
\\ 
\hline \hline 
$-1.10000$ & $ 3.4 \times 10^{-11} $ \\ 
$-1.09000$ & $ 3.8  \times 10^{-11} $ \\
$-1.08000$ & $ 4.3  \times 10^{-11} $\\
$-1.07000$ & $ 4.9  \times 10^{-11} $ \\
$-1.06000$ & $ 5.7  \times 10^{-11} $\\
$-1.05000$ & $ 6.9  \times 10^{-11} $\\
$-1.04000$ & $ 8.6  \times 10^{-11} $\\
$-1.03000$ & $ 1.1  \times 10^{-10} $  \\
$-1.02000$ & $ 1.7  \times 10^{-10} $\\
$-1.01000$ & $ 3.4  \times 10^{-10} $\\
$-1.00900$ & $ 3.8  \times 10^{-10} $\\
$-1.00800$ & $ 4.3  \times 10^{-10} $\\
$-1.00700$ & $ 4.9  \times 10^{-10} $\\
$-1.00600$ & $ 5.7  \times 10^{-10} $\\
$-1.00500$ & $ 6.9  \times 10^{-10} $\\
$-1.00400$ & $ 8.6  \times 10^{-10} $\\
$-1.00300$ & $ 1.1  \times 10^{-9} $\\
$-1.00200$ & $ 1.7  \times 10^{-9} $\\
$-1.00100$ & $ 3.4  \times 10^{-9} $ \\
$-1.00090$ & $ 3.8  \times 10^{-9} $\\
$-1.00080$ & $ 4.3  \times 10^{-9} $\\
$-1.00070$ & $ 4.9  \times 10^{-9} $\\
$-1.00060$ & $ 5.7  \times 10^{-9} $\\
$-1.00050$ & $ 6.9  \times 10^{-9} $ \\
$-1.00040$ & $ 8.6  \times 10^{-9} $ \\
$-1.00030$ & $ 1.1  \times 10^{-8} $\\
$-1.00020$ & $ 1.7  \times 10^{-8} $\\
$-1.00010$ & $ 3.4  \times 10^{-8} $\\
$-1.00009$ & $ 3.8  \times 10^{-8} $\\
$-1.00008$ & $ 4.3  \times 10^{-8} $\\
$-1.00007$ & $ 4.9  \times 10^{-8} $\\
$-1.00006$ & $ 5.7  \times 10^{-8} $\\
$-1.00005$ & $ 6.9  \times 10^{-8} $\\
$-1.00004$ & $ 8.6  \times 10^{-8} $\\
$-1.00003$ & $ 1.1  \times 10^{-7} $\\
$-1.00002$ & $ 1.7  \times 10^{-7} $\\
$-1.00001$ & $ 3.4  \times 10^{-7} $\\
\hline 
\end{tabular} 
\vspace{15pt} 
\caption{ Values of ($t_{\rm rip} - t_T$) for $-1.10000 \le \omega \le -1.00001$ with $\eta=10^{29}$ fixed.  } 
\label{triptotT29}
\end{center}

\end{table}

\begin{table} 

\begin{center}

$\eta=10^{57} $ 

\begin{tabular}{c|r} 
\hline 
\hspace{55pt} $w$  \hspace{55pt} & 
\hspace{55pt} $(t_{\rm rip} - t_T) $ [s] \hspace{55pt} 
\\ 
\hline \hline 
$-1.10000$ & $ 3.4 \times 10^{-25} $ \\ 
$-1.09000$ & $ 3.8  \times 10^{-25} $ \\
$-1.08000$ & $ 4.3  \times 10^{-25} $\\
$-1.07000$ & $ 4.9  \times 10^{-25} $ \\
$-1.06000$ & $ 5.7  \times 10^{-25} $\\
$-1.05000$ & $ 6.9  \times 10^{-25} $\\
$-1.04000$ & $ 8.6  \times 10^{-25} $\\
$-1.03000$ & $ 1.1  \times 10^{-24} $  \\
$-1.02000$ & $ 1.7  \times 10^{-24} $\\
$-1.01000$ & $ 3.4  \times 10^{-24} $\\
$-1.00900$ & $ 3.8  \times 10^{-24} $\\
$-1.00800$ & $ 4.3  \times 10^{-24} $\\
$-1.00700$ & $ 4.9  \times 10^{-24} $\\
$-1.00600$ & $ 5.7  \times 10^{-24} $\\
$-1.00500$ & $ 6.9  \times 10^{-24} $\\
$-1.00400$ & $ 8.6  \times 10^{-24} $\\
$-1.00300$ & $ 1.1  \times 10^{-23} $\\
$-1.00200$ & $ 1.7  \times 10^{-23} $\\
$-1.00100$ & $ 3.4  \times 10^{-23} $ \\
$-1.00090$ & $ 3.8  \times 10^{-23} $\\
$-1.00080$ & $ 4.3  \times 10^{-23} $\\
$-1.00070$ & $ 4.9  \times 10^{-23} $\\
$-1.00060$ & $ 5.7  \times 10^{-23} $\\
$-1.00050$ & $ 6.9  \times 10^{-23} $ \\
$-1.00040$ & $ 8.6  \times 10^{-23} $ \\
$-1.00030$ & $ 1.1  \times 10^{-22} $\\
$-1.00020$ & $ 1.7  \times 10^{-22} $\\
$-1.00010$ & $ 3.4  \times 10^{-22} $\\
$-1.00009$ & $ 3.8  \times 10^{-22} $\\
$-1.00008$ & $ 4.3  \times 10^{-22} $\\
$-1.00007$ & $ 4.9  \times 10^{-22} $\\
$-1.00006$ & $ 5.7  \times 10^{-22} $\\
$-1.00005$ & $ 6.9  \times 10^{-22} $\\
$-1.00004$ & $ 8.6  \times 10^{-22} $\\
$-1.00003$ & $ 1.1  \times 10^{-21} $\\
$-1.00002$ & $ 1.7  \times 10^{-21} $\\
$-1.00001$ & $ 3.4  \times 10^{-21} $\\
\hline 
\end{tabular} 
\vspace{15pt} 
\caption{ Values of ($t_{\rm rip} - t_T$) for $-1.10000 \le \omega \le -1.00001$ with $\eta=10^{57}$ fixed.  } 
\label{triptotT57}
\end{center}

\end{table}

\begin{table} 

\begin{center}

$\eta=10^{93} $ 

\begin{tabular}{c|r} 
\hline 
\hspace{55pt} $w$  \hspace{55pt} & 
\hspace{55pt} $(t_{\rm rip} - t_T) $ [s] \hspace{55pt} 
\\ 
\hline \hline 
$-1.10000$ & $ 3.4 \times 10^{-43} $ \\ 
$-1.09000$ & $ 3.8  \times 10^{-43} $ \\
$-1.08000$ & $ 4.3  \times 10^{-43} $\\
$-1.07000$ & $ 4.9  \times 10^{-43} $ \\
$-1.06000$ & $ 5.7  \times 10^{-43} $\\
$-1.05000$ & $ 6.9  \times 10^{-43} $\\
$-1.04000$ & $ 8.6  \times 10^{-43} $\\
$-1.03000$ & $ 1.1  \times 10^{-42} $  \\
$-1.02000$ & $ 1.7  \times 10^{-42} $\\
$-1.01000$ & $ 3.4  \times 10^{-42} $\\
$-1.00900$ & $ 3.8  \times 10^{-42} $\\
$-1.00800$ & $ 4.3  \times 10^{-42} $\\
$-1.00700$ & $ 4.9  \times 10^{-42} $\\
$-1.00600$ & $ 5.7  \times 10^{-42} $\\
$-1.00500$ & $ 6.9  \times 10^{-42} $\\
$-1.00400$ & $ 8.6  \times 10^{-42} $\\
$-1.00300$ & $ 1.1  \times 10^{-41} $\\
$-1.00200$ & $ 1.7  \times 10^{-41} $\\
$-1.00100$ & $ 3.4  \times 10^{-41} $ \\
$-1.00090$ & $ 3.8  \times 10^{-41} $\\
$-1.00080$ & $ 4.3  \times 10^{-41} $\\
$-1.00070$ & $ 4.9  \times 10^{-41} $\\
$-1.00060$ & $ 5.7  \times 10^{-41} $\\
$-1.00050$ & $ 6.9  \times 10^{-41} $ \\
$-1.00040$ & $ 8.6  \times 10^{-41} $ \\
$-1.00030$ & $ 1.1  \times 10^{-40} $\\
$-1.00020$ & $ 1.7  \times 10^{-40} $\\
$-1.00010$ & $ 3.4  \times 10^{-40} $\\
$-1.00009$ & $ 3.8  \times 10^{-40} $\\
$-1.00008$ & $ 4.3  \times 10^{-40} $\\
$-1.00007$ & $ 4.9  \times 10^{-40} $\\
$-1.00006$ & $ 5.7  \times 10^{-40} $\\
$-1.00005$ & $ 6.9  \times 10^{-40} $\\
$-1.00004$ & $ 8.6  \times 10^{-40} $\\
$-1.00003$ & $ 1.1  \times 10^{-39} $\\
$-1.00002$ & $ 1.7  \times 10^{-39} $\\
$-1.00001$ & $ 3.4  \times 10^{-39} $\\
\hline 
\end{tabular} 
\vspace{15pt} 
\caption{ Values of ($t_{\rm rip} - t_T$) for $-1.10000 \le \omega \le -1.00001$ with $\eta=10^{93}$ fixed.  } 
\label{triptotT93}
\end{center}

\end{table}

\newpage

\subsection*{A3. The formula for ($t_{\rm rip}-t_{\rm unbound}$)}

Let us next consider a time $t_{\rm unbound}$ at which 
point a gravitationally bound system will become unbound due to 
an extraordinarily rapid expansion of the universe. 
Roughly speaking, 
a bound system in circular orbit at radius $R$ with mass $M$
becomes unbound when 
\begin{equation}
  \frac{4\pi}{3} R^3  \rho_\Lambda(t_{\rm unbound})  |1+3 w| 
  \simeq M 
  \,, \label{unbound}
\end{equation}
where the left-hand side comes from the $T_{\mu\nu}$-term in 
the right-hand side of the Einstein equation.

Putting $t=t_{\rm unbound}$ in Eq.(\ref{tripto}) and 
rewriting the overall factor $(H_0  \sqrt{\Omega})^{-1}$,  
in terms of $\rho_\Lambda^0$ and the gravitational constant $G$, 
as $(H_0 \sqrt{\Omega})^{-1}= (8\pi G/3 \cdot \rho_\Lambda^0)^{-1/2}$, 
we may express a time interval $(t_{\rm rip} - t_{\rm unbound})$ as  
\begin{eqnarray} 
  (t_{\rm rip} -t_{\rm unbound}) 
&=& (8\pi G/3  \rho_\Lambda^0)^{-1/2}
 \frac{2}{3|1+w|} [a(t_{\rm unbound})]^{-\frac{3|1+w|}{2}} 
\nonumber\\ 
&=&
(8\pi G/3)^{-1/2}  \sqrt{\frac{1}{\rho_\Lambda^0}}
 \frac{2}{3|1+w|} \sqrt{ \frac{\rho_\Lambda^0 }{\rho_\Lambda(t_{\rm unbound})} }
\nonumber \\ 
&=&
(8\pi G/3)^{-1/2} 
 \frac{2}{3|1+w|} \sqrt{ \frac{1}{\rho_\Lambda(t_{\rm unbound})} }
\,. \label{trip:tunbound}
\end{eqnarray}
  Using Eq.(\ref{unbound}) we can further rewrite the right-hand side as 
\begin{eqnarray} 
  (t_{\rm rip} -t_{\rm unbound})  
&=& 
\frac{\sqrt{2|1+3w|}}{3|1+w|} 
 \sqrt{ \frac{R^3}{G M}} 
\nonumber \\ 
&=& 
\frac{\sqrt{2|1+3w|} }{6\pi |1+w|} 
 P
\,, 
\end{eqnarray}  
where in the last line we have used a relationship from classical gravitational systems, 
\begin{equation} 
 P=2\pi \sqrt{\frac{R^3}{GM}} 
 \,, 
\end{equation} 
in which $P$ denotes the period for a circular orbit of radius $R$ around a system bound by gravitational force 
with mass $M$. 
Thus we reach the expression~\cite{Caldwell:2003vq} for the time interval $( t_{\rm rip} - t_{\rm unbound})$, 
\begin{equation} 
  (t_{\rm rip} - t_{\rm unbound}) = \alpha(w)  P 
  \,, \label{tueq}
\end{equation}
where  
\begin{equation} 
 \alpha (w) = \frac{\sqrt{2 |1+3w|}}{6\pi |1+w|} 
 \,. 
\end{equation}

Similarly to Eq.(\ref{unbound}), 
even for binding forces other than gravity, 
we can roughly estimate an unbound-time $t_{\rm unbound}$.  
 For simplicity, we shall derive here a relationship similar to Eq.(\ref{unbound}) 
focusing on an electromagnetically bound system, 
e.g., a hydrogen atom $H$, 
in which the electron is 
constrained on a circular orbit of radius $R$ by the Coulomb force 
around a proton. 
This can be done just by taking into account 
the balance problem between the Coulomb force $F_C$ and the dark energy force $F_w$. 
We find that 
the system will become unbound when    
\begin{eqnarray}
  F_C &\simeq &  F_w(t_{\rm unbound}) 
  \,, \nonumber \\ 
  \to \hspace{20pt}
  \frac{e^2}{4\pi \epsilon_0} \frac{1}{R^2} 
  &\simeq & G \frac{ m_e m_{\rm eff}(w, t_{\rm unbound}) }{R^2} 
  \,, \label{balance:em}
\end{eqnarray}
where $m_{\rm eff}(w, t_{\rm unbound})$ denotes an effective ``mass" arising from 
the dark energy density at $t=t_{\rm unbound}$, 
\begin{equation} 
 m_{\rm eff}(w, t_{\rm unbound}) 
 = 
 \frac{4\pi R^3}{3}  \rho_\Lambda(t_{\rm unbound})  |1+3 w| 
 \,. 
\end{equation}
Using the expression for the period associated with the electromagnetic force,  
\begin{equation} 
  P_{\rm em} = 2 \pi \sqrt{ \frac{m_e R^3}{\hbar c \alpha} }, 
\end{equation}
we can rewrite Eq.(\ref{balance:em}) as 
\begin{equation} 
  \rho_\Lambda(t_{\rm unbound})  |1+3 w| 
  \simeq 
  \frac{3\pi}{ G P_{\rm em}^2} 
  \,, 
\end{equation}
which leads immediately to the formula for $(t_{\rm rip}-t_{\rm unbound}^{\rm H})$ in the case of an H atom. 
As a result, we find  it take the same form as Eq.(\ref{tueq}), 
\begin{equation} 
  t_{\rm rip} - t_{\rm unbound}^{\rm H} = \alpha(w)  P_{\rm em} 
  \,. \label{tueq:H}
\end{equation}

 It is straightforward to show that, for other binding forces (e.g. strong forces, etc.), 
the form of Eq.(\ref{tueq:H}) is unchanged except for replacing $P_{\rm em}$ with 
the appropriate one associated with the binding force.

   Choosing typical bound systems -- galaxy, Sun-Earth, and hydrogen atom -- 
and supplying the corresponding values for the period $P(P_{\rm em})$ 
\begin{center}
\begin{tabular}{c|r|r} 
\hline 
\hspace{30pt} Bound System  \hspace{30pt} & 
\hspace{30pt} $P$ \hspace{30pt} & 
\hspace{30pt} $L$ \hspace{30pt} 
\\ 
\hline \hline 
Typical Galaxy & $2.0 \times 10$ yr & $1.6 \times 10^4$ pc \\ 
Sun-Earth &  1 yr & $1.5 \times 10^8$ km \\ 
Hydrogen Atom & $10^{-16}$ s & $0.5 \times 10^{-10}$ m \\ 
\hline 
\end{tabular} 
\end{center} 
we calculate the values of $( t_{\rm rip} - t_{\rm unbound})$ for each bound system 
by taking values of $w$ from the range $-1.10000 \le \omega \le -1.00001$. 
The result is summarized in Table~\ref{triptotU}.

  \begin{table} 

\begin{center}

( $t_{\rm rip} - t_{\rm unbound}) $

\begin{tabular}{c|r|r|r} 
\hline 
\hspace{5pt} $w$  \hspace{5pt} & 
\hspace{5pt} \footnotesize{Typical Galaxy [Gyr]  } \hspace{5pt} & 
\hspace{5pt} \footnotesize{Sun-Earth [yr]   } \hspace{5pt} & 
\hspace{5pt} \footnotesize{Hydrogen Atom [s] }  \hspace{5pt} 
\\ 
\hline \hline 
$-1.10000$ & 0.22 & 1.13 & $1.13 \times 10^{-16}$  \\ 
$-1.09000$ & 0.25 & 1.25 & $1.25 \times 10^{-16}$ \\ 
$-1.08000$ & 0.28 & 1.40 & $1.40 \times 10^{-16}$ \\ 
$-1.07000$ & 0.31 & 1.59 & $1.59 \times 10^{-16}$ \\ 
$-1.06000$ & 0.36 & 1.84 & $1.84 \times 10^{-16}$ \\ 
$-1.05000$ & 0.44 & 2.20 & $2.20 \times 10^{-16}$ \\ 
$-1.04000$ & 0.54 & 2.73 & $2.73 \times 10^{-16}$ \\ 
$-1.03000$ & 0.72 & 3.61 & $3.61 \times 10^{-16}$ \\ 
$-1.02000$ & 1.07 & 5.38 & $5.38 \times 10^{-16}$ \\ 
$-1.01000$ & 2.13 & 10.6 & $1.06 \times 10^{-15}$ \\ 
$-1.00900$ & 2.37 & 11.8 & $1.18 \times 10^{-15}$ \\ 
$-1.00800$ & 2.66 & 13.3 & $1.33 \times 10^{-15}$ \\ 
$-1.00700$ & 3.04 & 15.2 & $1.52 \times 10^{-15}$ \\ 
$-1.00600$ & 3.55 & 17.7 & $1.77 \times 10^{-15}$ \\ 
$-1.00500$ & 4.26 & 21.3 & $2.13 \times 10^{-15}$ \\ 
$-1.00400$ & 5.32 & 26.6 & $2.66 \times 10^{-15}$ \\ 
$-1.00300$ & 7.08 & 35.4 & $3.54 \times 10^{-15}$ \\ 
$-1.00200$ & 10.2 & 53.1 & $5.31 \times 10^{-15}$ \\ 
$-1.00100$ & 21.2 & 106 & $1.06 \times 10^{-14}$ \\ 
$-1.00090$ & 23.5 & 117 & $1.17 \times 10^{-14}$ \\ 
$-1.00080$ & 26.5 & 132 & $1.32 \times 10^{-14}$ \\ 
$-1.00070$ & 30.3 & 151 & $1.51 \times 10^{-14}$ \\ 
$-1.00060$ & 35.3 & 176 & $1.76 \times 10^{-14}$ \\ 
$-1.00050$ & 42.4 & 212 & $2.12 \times 10^{-14}$ \\ 
$-1.00040$ & 53.0 & 265 & $2.65 \times 10^{-14}$ \\ 
$-1.00030$ & 70.7 & 353 & $3.53 \times 10^{-14}$ \\ 
$-1.00020$ & 106 & 530 & $5.30 \times 10^{-14}$ \\ 
$-1.00010$ & 212 & 1061 & $1.06 \times 10^{-13}$ \\ 
$-1.00009$ & 235 & 1179 & $1.17 \times 10^{-13}$ \\ 
$-1.00008$ & 265 & 1326 & $1.32 \times 10^{-13}$ \\ 
$-1.00007$ & 303 & 1515 & $1.51 \times 10^{-13}$ \\ 
$-1.00006$ & 353 & 1768 & $1.76 \times 10^{-13}$\\ 
$-1.00005$ & 424 & 2122 & $2.12 \times 10^{-13}$ \\ 
$-1.00004$ & 530 & 2652 & $2.65 \times 10^{-13}$ \\ 
$-1.00003$ & 707 & 3536 & $3.53 \times 10^{-13}$ \\ 
$-1.00002$ & 1060 & 5305 & $5.30 \times 10^{-13}$ \\ 
$-1.00001$ & 2120 & 10610 & $1.06 \times 10^{-12}$ \\ 
\hline 
\end{tabular} 
\vspace{15pt} 
\caption{ Values of ($t_{\rm rip} - t_{\rm unbound}$) for $-1.10000 \le \omega \le -1.00001$  } 
\label{triptotU}
\end{center}

\end{table}

\newpage

\subsection*{A4. The formula for ($t_{\rm rip}-t_{\rm caus}$)}

After $t=t_{\rm unbound}$, 
objects which had been constrained in  a bound system 
would be free to move far apart and will end up causally disconnected starting at time $t=t_{\rm caus}$. 
Such a time, $t_{\rm caus}$, can be defined, with $c=1$ taken, by 
\begin{equation} 
 \frac{L}{a(t_{\rm caus})} 
 = \int_{t_{\rm caus}}^{t_{\rm rip}} \frac{dt}{a(t)} 
\,,  \label{comv}
\end{equation}
where $L$ denotes the length scale at which two objects separate at $t=t_{\rm caus}$, 
and the right-hand side stands for the comoving distance of light which arises from traveling 
at light speed $c$ during a time-interval $t_{\rm caus}<t<t_{\rm rip}$. 
Noting that $dt = da/(aH)$ and rewriting $H(a)$ 
from the Friedman equation in terms of a function of $a$, 
we calculate more explicitly 
the right-hand side of Eq.(\ref{comv}) as follows: 
\begin{eqnarray} 
   \frac{L}{a(t_{\rm caus})} 
 &=& 
  \int_{a(t_{\rm caus})}^{a(t_{\rm rip})} \frac{ da}{a^2 H(a)} 
\nonumber\\ 
 &=& 
(H_0 \cdot \Omega_{\Lambda}^0 )^{-1}
\int_{a(t_{\rm caus})}^{\infty} da \, a^{-1/2(1-3w)}  
\nonumber\\ 
 &=& 
(H_0  \Omega_{\Lambda}^0 )^{-1}
\frac{2}{|1+3w|}  [a(t_{\rm caus}) ]^{-|1+3w|/2}  
\,. \label{caus2}
\end{eqnarray}

 Taking $t=t_{\rm caus}$ in Eq.(\ref{tripto}) and dividing  
both sides by the resultant expression, 
we can continue calculating to get 
\begin{eqnarray} 
\frac{L}{a(t_{\rm caus})}  \frac{1}{( t_{\rm rip} - t_{\rm caus} )} 
&=&  \frac{3|1+w|}{|1+3w|} 
\left( \frac{[a(t_{\rm caus})]^{3|1+w|}}{[a(t_{\rm caus})]^{|1+3w|}} \right)^{1/2} 
\nonumber \\ 
&=&   \frac{3|1+w|}{|1+3w|}  \frac{1}{a(t_{\rm caus})} 
\,, 
\end{eqnarray} 
 and in the end we reach the expression~\cite{Caldwell:2003vq} 
\begin{equation} 
  t_{\rm rip} - t_{\rm caus} 
= \Bigg|  \frac{1+3w}{3(1+w)}  \Bigg| \frac{L}{c}
\,.\label{triptotcaus:formula}
\end{equation}
  Similarly to the previous section, 
we take values of $w$ in the range $-1.10000 \le \omega \le -1.00001$ and 
calculate the values of $( t_{\rm rip} - t_{\rm caus} )$ for 
typical bound systems such as galaxies, Sun-Earth, hydrogen atom,  
and summarize in Table~\ref{triptotcaus}.


\begin{table} 

\begin{center}

($t_{\rm rip} - t_{\rm caus}$)

\begin{tabular}{c|r|r|r} 
\hline 
\hspace{5pt} $w$  \hspace{5pt} & 
\hspace{5pt} \footnotesize{Typical Galaxy [Myr]  } \hspace{5pt} & 
\hspace{5pt} \footnotesize{Sun-Earth [day]   } \hspace{5pt} & 
\hspace{5pt} \footnotesize{Hydrogen Atom [s] }  \hspace{5pt} 
\\ 
\hline \hline 
$-1.10000$ & 0.40 & 0.044 & $1.27 \times 10^{-18}$  \\ 
$-1.09000$ & 0.44 & 0.048 & $1.40 \times 10^{-18}$ \\
$-1.08000$ & 0.49 & 0.054 & $1.55 \times 10^{-18}$ \\
$-1.07000$ & 0.55 & 0.060 & $1.75 \times 10^{-18}$ \\
$-1.06000$ & 0.64 & 0.070 & $2.01 \times 10^{-18}$ \\
$-1.05000$ & 0.75 & 0.083 & $2.39 \times 10^{-18}$ \\
$-1.04000$ & 0.93 & 0.102 & $2.94 \times 10^{-18}$ \\
$-1.03000$ & 1.22 & 0.134 & $3.87 \times 10^{-18}$ \\
$-1.02000$ & 1.81 & 0.198 & $5.72 \times 10^{-18}$ \\
$-1.01000$ & 3.57 & 0.391 & $1.12 \times 10^{-17}$ \\
$-1.00900$ & 3.96 & 0.434 & $1.25\times 10^{-17}$ \\
$-1.00800$ & 4.45 & 0.488 & $1.40 \times 10^{-17}$ \\
$-1.00700$ & 5.08 & 0.557 & $1.60 \times 10^{-17}$ \\
$-1.00600$ & 5.92 & 0.649 & $1.86 \times 10^{-17}$ \\
$-1.00500$ & 7.09 & 0.777 & $2.24 \times 10^{-17}$ \\
$-1.00400$ & 8.86 & 0.970 & $2.79 \times 10^{-17}$ \\
$-1.00300$ & 11.7 & 1.29   & $3.72 \times 10^{-17}$ \\
$-1.00200$ & 17.6 & 1.93   & $5.57 \times 10^{-17}$ \\
$-1.00100$ & 35.2 & 3.86   & $1.11 \times 10^{-16}$ \\
$-1.00090$ & 39.2 & 4.29   & $1.23 \times 10^{-16}$ \\
$-1.00080$ & 44.0 & 4.83   & $1.39 \times 10^{-16}$ \\
$-1.00070$ & 50.3 & 5.52   & $1.59 \times 10^{-16}$ \\
$-1.00060$ & 58.7 & 6.44   & $1.85 \times 10^{-16}$ \\
$-1.00050$ & 70.5 & 7.72   & $2.22 \times 10^{-16}$ \\
$-1.00040$ & 88.1 & 9.65   & $2.78 \times 10^{-16}$ \\
$-1.00030$ & 117 & 12.8    & $3.70 \times 10^{-16}$ \\
$-1.00020$ & 176 & 19.3    & $5.56 \times 10^{-16}$ \\
$-1.00010$ & 352 & 38.6    & $1.11 \times 10^{-15}$ \\
$-1.00009$ & 391 & 42.9    & $1.23 \times 10^{-15}$ \\
$-1.00008$ & 440 & 48.2    & $1.39 \times 10^{-15}$ \\
$-1.00007$ & 503 & 55.1    & $1.58 \times 10^{-15}$ \\
$-1.00006$ & 587 & 64.3    & $1.85 \times 10^{-15}$ \\
$-1.00005$ & 704 & 77.2    & $2.22 \times 10^{-15}$ \\ 
$-1.00004$ & 880 & 96.5    & $2.77 \times 10^{-15}$ \\
$-1.00003$ & 1170 & 128   & $3.70 \times 10^{-15}$ \\
$-1.00002$ & 1760 & 193   & $5.55 \times 10^{-15}$ \\
$-1.00001$ & 3520 & 386   & $1.11 \times 10^{-14}$ \\
\hline 
\end{tabular} 
\vspace{15pt} 
\caption{ Values of ($t_{\rm rip} - t_{\rm caus}$) for $-1.10000 \le \omega \le -1.00001$  } 
\label{triptotcaus}
\end{center}

\end{table}

\newpage

\Large 

\noindent 
{\bf Appendix B}

\bigskip 

\noindent 
{\bf Constraint on $\rho_C$ from causality}  

\bigskip 

\normalsize

There exists a lower bound on the value of $\rho_C$ 
coming from the causal disconnection condition 
for smaller bound systems such as the hydrogen atom and nucleon. 
We study here the lower bound on $\rho_C$ 
calculating $(t_T - t_{\rm caus})$ numerically as a function 
of $w$ and $\rho_C$ for each bound system with a size$\lesssim 10^{-10}$m.

\begin{figure}
\begin{center}
\vspace{15pt}
\psfrag{y}{\vspace{40pt} \hspace{-45pt} $-\log_{10}|1+w|$}
\psfrag{x}{\vspace{-20pt} \hspace{-25pt} $\log_{10} \eta$  } 
\vspace{15pt} 
\includegraphics[scale=0.8]{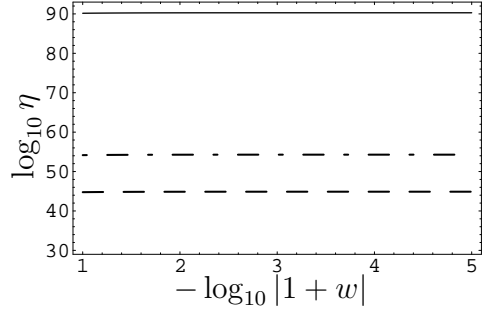} 
\vspace{5pt}
\caption{\footnotesize The lower bound on $\eta$ coming from the causal disconnection conditions, 
$t_T>t^{\rm H}_{\rm caus}$ (dashed line), 
$t_T>t^{\rm N}_{\rm caus}$ (dashed-dotted line), 
$t_T>t^{\rm PPP}_{\rm caus}$ (solid line), 
for $10^{29} \le \eta \le 10^{93}$ and $-1.10000 \le w \le -1.00001$.  
The regions below these three lines are forbidden by causality. }  
\label{cons-rhoc-HNPPP} 
\end{center}

\end{figure}

Imposing the condition $t_T>t_{\rm caus}^{\rm H, N}$ where $H, N$ denote the hydrogen atom and nucleon, respectively, 
we may obtain physical constraints on $\eta$ and $w$ from the following inequality: 
\begin{eqnarray} 
(t_T - t_{\rm caus})^{\rm H, N} 
&=& 
\frac{|1+3w| (L_{\rm H,N}/c) }{3|1+w|}
\left[ 1 - \frac{3.12 \times 10^{11.95}/L_{\rm H, N}}{\sqrt{\eta} |1+3w| } \right]
\nonumber \\  
 & \ge &0 
  \,, \label{causdisco-condi}
\end{eqnarray} 
which follows from Eqs.(\ref{triptotT}) and (\ref{triptotcaus:formula}). 
In Fig~\ref{cons-rhoc-HNPPP} we show the constraints on the model 
parameters $\eta$ and $w$, coming from 
the causal disconnection condition (\ref{causdisco-condi}). 
Here we have taken 
$L_{\rm H}=0.5 \times 10^{-10}$m, $L_{\rm N}=10^{-15}$m, 
and \cite{Yao:2006px} $c=2.9979 \times 10^{8}\,{\rm m} \, {\rm s}^{-1}$, 
yr = $3.1556 \times 10^7$ s. 
  From this figure, 
we find a lower bound  on the value of $\eta$, 
or equivalently on the critical density $\rho_C$, 
\begin{equation} 
   \eta \gtrsim 10^{54}
   \quad 
   \leftrightarrow 
   \quad 
      \rho_C \gtrsim (10^9 \, {\rm GeV})^4 
\,, 
\end{equation}
where we have calculated $\rho_C = (1.44 \times 10^{x/4-4.5} \, {\rm GeV})^4$ with $x=\log_{10} \eta$.

If we extend a similar study to a bound PPP system, 
for which we set a scale $L_{\rm PPP}\simeq 10^{-33}$m,
slightly above the Planck scale,  
we find a stronger lower bound on $\eta$ (see the solid line illustrated in Fig.~\ref{cons-rhoc-HNPPP}), 
\begin{equation} 
   \eta \gtrsim 10^{90}
   \quad 
   \leftrightarrow 
   \quad 
      \rho_C \gtrsim (10^{18} \, {\rm GeV})^4 
\,. 
\end{equation}

It is interesting to note that the result on 
these lower bounds is fairly insensitive to the choice of 
$w$ within the range of interest.

\newpage

\newpage 

.

\newpage

\Large 

\noindent 
{\bf Appendix C}

\bigskip 

\noindent 
{\bf Comes back empty condition}  

\bigskip 

\normalsize

At $t=t_T$ we require that at deflation immediately
prior to turnaround of the cyclic universe 
the causal patch {\it comes back  
empty} which demands that the deflation factor $f$ satisfies $f^{-3} \gtrsim 10^{102}$
in order to solve the entropy problem, since the present entropy
is at least $10^{102}$~\cite{FK,FHKR}.

 This requirement can be met by imposing 
 that one causal patch 
be less than a size of the smallest bound systems, $L_p(t_T)$
\begin{equation} 
  \frac{r_H(t_T)}{N_{\rm cp}} \le L_p(t_T) 
  \,, \label{cbe1}
\end{equation}
where the subscript $p$ denotes a bound system whose size $L_0$ at present
lies in the range $10^{-33}$m $\le L_0 \le 10^{-15}$m. 
Noting that 
\begin{eqnarray}
  r_H(t_T) &=& r_{H_0}  a(t_T) \,, 
  \\ 
  L_p(t_T) &=& L_p(t_0) \frac{a(t_T)}{a(t_{\rm unbound})}
  \,, 
\end{eqnarray}
we can rewrite the condition (\ref{cbe1}) as 
\begin{equation}
 a(t_{\rm unbound}) 
 \le N_{\rm cp}  \frac{L_p(t_0)}{r_{H_0}} 
 \,. \label{cbe2} 
\end{equation}
Looking at Eqs.(\ref{trip:tunbound}), (\ref{tueq}) and (\ref{tueq:H}), 
we see that the left hand side of Eq.(\ref{cbe2}) can be reexpressed as 
\begin{eqnarray} 
  \left[ \frac{\sqrt{2 |1 + 3w| }}{4 \pi} P_p  (H_0  \sqrt{\Omega_{\Lambda}^0}) \right]^{\frac{2}{-3|1+w|}} 
  \le N_{\rm cp}  \frac{L_p(t_0)}{r_{H_0}} 
  \,, \label{cbe3}
\end{eqnarray}
where period $P_p$, associated with a certain smaller bound system $p$ we are concerned with,  
can be expressed as  $P_p \simeq L_p(t_0)/c$. 

Taking $\log_{10}\log_{10}$ of $N_{\rm cp}$
makes it easier to plot the inequality (\ref{cbe3}). 
A plot of the ($w$, $\log_{10}\log_{10}N_{\rm cp}$)-plane 
varying the value of $L_p(t_0)$ in the range of interest,  
$10^{-33}$m $\le L_p(t_0) \le 10^{-15}$m  
is shown in Figure~\ref{cons-w-cbe}.  
We have used~\cite{Yao:2006px}  
$r_{H_0}=1.232 \times 10^{26}$m 
and 
$\Omega_\Lambda^0  = 0.76$.

\end{document}